\documentclass[a4paper]{article}

\usepackage{INTERSPEECH2021}
\usepackage{amsmath,amsfonts,amssymb,mathtools}
\usepackage{booktabs}
\usepackage{multirow}
\usepackage{caption}
\usepackage{enumitem}
\usepackage{hyperref}
\usepackage{url}
\usepackage{tikz}
\usepackage{xfrac}
\usetikzlibrary{decorations.text}
\usetikzlibrary{arrows.meta}
\usepackage{pgfplots}
\usepackage{array}
\usepackage{siunitx}
\usepackage{wrapfig}
\usepackage{cite}
\pgfplotsset{compat=1.16}

\title{WaveGrad 2: Iterative Refinement for Text-to-Speech Synthesis}
\name{Nanxin~Chen\thanks{$^\dagger$Work done during an internship at Google Brain.}$^{1\dagger}$, Yu~Zhang$^2$, ~~Heiga~Zen$^2$,~~Ron~J.~Weiss$^2$,~~Mohammad~Norouzi$^2$,~~Najim Dehak$^1$, ~~William Chan$^2$}
\address{
  $^1$Center for Language and Speech Processing, Johns Hopkins University\\
  $^2$Brain Team, Google Research}
\email{\{nchen14,ndehak3\}@jhu.edu,\{ngyuzh,heigazen,ronw,mnorouzi,williamchan\}@google.com}

\begin{document}

\maketitle
\begin{abstract}
This paper introduces \textit{WaveGrad~2}, a non-autoregressive generative model for text-to-speech synthesis.
WaveGrad~2 is trained to estimate the gradient of the log conditional density of the waveform given a phoneme sequence.
The model takes an input phoneme sequence, and through an iterative refinement process, generates an audio waveform. This contrasts to the original WaveGrad vocoder which conditions on mel-spectrogram features, generated by a separate model. The iterative refinement process starts from Gaussian noise, and through a series of refinement steps (e.g., 50 steps), progressively recovers the audio sequence. 
WaveGrad~2 offers a natural way to trade-off between inference speed and sample quality, through adjusting the number of refinement steps.
Experiments show that the model can generate high fidelity audio, approaching the performance of a state-of-the-art neural TTS system.
We also report various ablation studies over different model configurations.
Audio samples are available at {\small \url{https://wavegrad.github.io/v2}}.

\end{abstract}

\section{Introduction}
Deep learning has revolutionized text-to-speech (TTS) synthesis~\cite{vanwavenet,char2wav,oord2018parallel,tacotron,shen2018natural,li-aaai-2019,ren2020fastspeech}.
Text-to-speech is a multimodal generation problem, which maps a input text sequence to a speech sequence with many possible variations in, for example, prosody, speaking style, and phonation modes.
Most neural TTS systems follow a two-stage generation process.
In the first step, a \textit{feature generation model} generates intermediate representations, typically linear or mel-spectrograms, from a text or phoneme sequence.
The intermediate representations control the structure of the waveform and are usually generated by an autoregressive architecture~\cite{tacotron,shen2018natural,li-aaai-2019,yu2019durian,shen2020non} to capture rich distributions.
Next, a \textit{vocoder} takes the intermediate features as input and predicts the waveform \cite{shen2018natural,prenger2019waveglow, ping2018clarinet, kim2019flowavenet, kim2020wavenode, wu2020waveffjord,donahue2018adversarial, engel2018gansynth, kumar-neurips-2019, yang-arxiv-2020, yamamoto-icassp-2020, binkowski-iclr-2020, yang2020vocgan, mccarthy2020hooligan,chen2021wavegrad}.  Because it takes the predicted features from the feature generation model as input during inference, the vocoder is often trained using predicted features as input~\cite{shen2018natural}.

Even though this two-stage TTS pipeline can produce high-fidelity audio, the deployment can be complicated since it uses a cascade of learned modules.
Another concern is related to the intermediate features which are often chosen largely by experience.
For example, mel-spectrogram features usually work well but may not be the best choice for all applications.
In contrast, the benefits of data-driven end-to-end approaches have been widely observed in various task domains across machine learning.
End-to-end approaches are able to learn the best intermediate features automatically from the training data, which are usually task-specific.
They are also easier to train since they do not require supervision and ground truth signal at different stages.

There are two predominant end-to-end TTS models.
\textit{Autoregressive} models offer tractable likelihood computation, but they require recurrent generation of waveform samples at inference time~\cite{char2wav}, which can be slow. %
By contrast, \textit{non-autoregressive} models enable efficient parallel generation, but they require token duration information.
Reference token durations for training are usually computed by an offline forced-alignment model~\cite{yu2019durian,ren2020fastspeech}.
To predict durations for generation, an additional module is trained to predict the ground truth durations.
More recent work~\cite{donahue2020end,ren2020fastspeech} has focused on applying non-autoregressive models to end-to-end TTS.
However, they still rely on spectral losses and mel-spectrograms for alignment and do not take full advantage of end-to-end training.
FastSpeech~2~\cite{ren2020fastspeech} requires additional conditioning signals such as pitch and energy to reduce the number of candidate output sequences.
EATS~\cite{donahue2020end} uses adversarial training as well as spectrogram losses to handle the one-to-many mapping issue, which makes the architecture more complicated.

In this work, we propose WaveGrad~2, a non-autoregressive phoneme-to-waveform model that does not require intermediate features or specialized loss functions.
To make the architecture more end-to-end, the WaveGrad~\cite{chen2021wavegrad} decoder is integrated into a Tacotron~2-style non-autoregressive encoder.
The WaveGrad decoder iteratively refines the input signal, beginning from random noise, and can produce high-fidelity audio with sufficient steps.
The one-to-many mapping problem is handled by the score matching objective, which optimizes weighted variational lower-bound of the log-likelihood~\cite{ho-arxiv-2020}.

The non-autoregressive encoder follows recently proposed non-attentive Tacotron~\cite{shen2020non} that combines a text encoder and a Gaussian resampling layer to incorporate the duration information.
The ground-truth duration information is utilized during training and a duration predictor is trained to estimate it.
During inference, the duration predictor predicts the duration for each input token.
Compared to the attention-based models, such a duration predictor is significantly more resilient to attention failures and monotonic alignments are guaranteed because of the way to compute the position.
The main contributions of this paper are as follows:
\begin{itemize}[noitemsep,leftmargin=*]
    \item A fully differentiable and efficient architecture that produces waveforms directly without generating intermediate features like spectrograms explicitly;
    \item A model providing a natural trade-off between fidelity and generation speed by changing the number of refinement steps;
    \item An end-to-end non-autoregressive model reaches 4.43 Mean Opinion Score (MOS), nearing performance of state-of-the-art neural TTS systems. %
\end{itemize}

\section{Score matching}\label{sec:scorematching}
Similar to the original WaveGrad~\cite{chen2021wavegrad}, WaveGrad~2 is built on prior work on score matching~\cite{hyvarinen2005estimation,vincent-neco-2011} and diffusion probabilistic models~\cite{sohl2015deep,ho-arxiv-2020}.
In the case of TTS, the score function is defined as the gradient of the log conditional distribution $p(y \mid x)$ with respect to the output $y$ as
\begin{equation}
    s(y \mid x) = \nabla_y \log p(y \mid x),
\end{equation}
where $y$ is the waveform and $x$ is the conditioning signal.
To synthesize speech given the conditional signal, one can draw a waveform iteratively via Langevin dynamics starting from the initialization $\tilde y_0$ as
\begin{equation}
    \tilde y_{i+1} = \tilde y_i + \frac{\eta}{2} s(\tilde y_i \mid x) + \sqrt{\eta}\, z_i,
\end{equation}
where $\eta > 0$ is the step size, $z_i \sim \mathcal{N}(0, I)$, and $I$ denotes an identity matrix.

Following previous work~\cite{chen2021wavegrad}, we adopt a special parameterization known as the diffusion model~\cite{sohl2015deep,ho-arxiv-2020}. A score network $s(\tilde y \mid x, \bar \alpha)$ is trained to predict the scaled derivative by minimizing the distance between model prediction and ground truth $\epsilon$ as
\begin{align}
     \mathbb{E}_{\bar \alpha,\epsilon}\left[ \left\lVert \epsilon_\theta\left(\tilde y, x, \sqrt {\bar \alpha} \right) - \epsilon \right\rVert_1 \right], \label{eqn:wavegradloss}
\end{align}
where $\epsilon \sim \mathcal{N}(0, I)$ is the noise term introduced by applying the reparameterization trick, $\bar\alpha$ is the noise level and $\tilde y$ is sampled according to
\begin{equation}
    \tilde y = \sqrt{\bar \alpha}\, y_0 +\sqrt{1 - \bar\alpha}\, \epsilon.
\end{equation}
During training, $\bar\alpha$'s are sampled from the intervals $\left[ \bar \alpha_n, \bar \alpha_{n+1} \right]$ based on a pre-defined linear schedule of $\beta$'s, according to:
\begin{equation}
\bar \alpha_n \coloneqq \prod_{s=1}^n (1 - \beta_s).
\end{equation}
In each iteration, the updated waveform is estimated following the following stochastic process
\begin{equation}
     y_{n-1} = \frac{1}{\sqrt{\alpha_n}}\left( y_n - \frac{\beta_n}{\sqrt{1-\bar\alpha_n}}\, \epsilon_\theta(y_n, x, \sqrt{\bar\alpha_n}) \right) + \sigma_n z.
\label{eq:inf}
\end{equation}

\section{WaveGrad 2}
\begin{figure}%
  \centering
  \resizebox{0.8\columnwidth}{!}{\begin{tikzpicture}[node distance=1cm, scale=1.0, every node/.style={transform shape}]
\tikzstyle{layer} = [rectangle, thick, minimum width=3.0cm, minimum height=0.6cm, align=center, draw=black]
\tikzstyle{block} = [rectangle, thick, rounded corners, minimum width=3.0cm, minimum height=0.6cm, align=center, draw=black]
\tikzstyle{block_small} = [rectangle, thick, rounded corners, minimum height=0.6cm, align=center, draw=black]
\tikzstyle{arrow} = [thick,->,-{Latex[length=2.5mm,width=2.5mm]}]
\tikzstyle{darrow} = [thick,->,-{Latex[length=2.5mm,width=2.5mm]},dashed]
\tikzstyle{line} = [thick]

\node(i0) [layer, fill=blue!20] at (2, 0) {Phoneme};
\node(l0) [block, fill=red!10!blue!10!] at (2, 1.2) {Encoder};
\node(i1) [layer, fill=blue!20] at (-2, 2.4) {Phoneme Duration};
\node(o0) [align=center] at (5.5, 2.4) {Duration Loss};
\node(l1) [block, fill=yellow!10!blue!10!] at (2, 2.4) {Duration Predictor};
\node(l2) [block, fill=yellow!30!blue!10!] at (2, 3.6) {Resampling};
\node(l3) [block, fill=blue!10!yellow!30!] at (2, 4.8) {Sampling window};
\node(l4) [block, fill=yellow!20!] at (2, 6.0) {WaveGrad Decoder};
\node(o1) [align=center] at (2, 7.2) {L1 loss};

\draw[arrow] (i0.north) -- (l0.south);
\draw[arrow] (l0.north) -- (l1.south);
\draw[arrow] (l1.north) -- (l2.south);
\draw[arrow] (l2.north) -- (l3.south);
\draw[arrow] (l3.north) -- (l4.south);
\draw[darrow] (l4.north) -- (o1.south);
\draw[darrow] (i1.east) -- (l1.west);
\draw[darrow] (l1.east) -- (o0.west);

\end{tikzpicture}}
  \caption{WaveGrad 2 network architecture. The inputs consists of the phoneme sequence. Dashed lines indicates computation only performed during training.}
  \label{fig:architecture}
\end{figure}
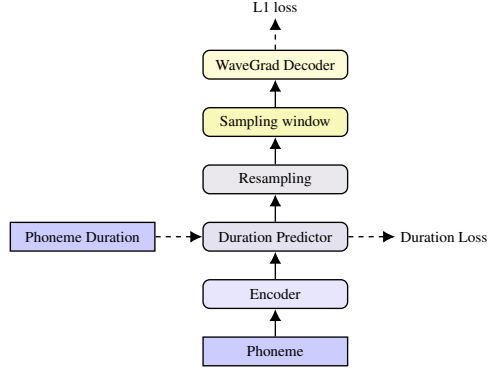

The proposed model includes three modules illustrated in Figure~\ref{fig:architecture}:
\begin{itemize}[leftmargin=*]
    \item The encoder takes a phoneme sequence as input and extracts abstract hidden representations from the input context.
    \item The resampling layer changes the resolution of the encoder output to match the output waveform time scale, quantized into 10ms segments (similar to typical mel-spectrogram features). This is achieved by conditioning on the target duration during training.  Durations predicted by the duration predictor module are utilized during inference.
    \item The WaveGrad decoder predicts the raw waveform by refining the noisy waveform iteratively. In each iteration, the decoder gradually refines the signal and adds fine-grained details.
\end{itemize}

\subsection{Encoder}

The design of the encoder follows that of Tacotron~2~\cite{shen2018natural}.
Phoneme tokens are used as inputs, with silence tokens inserted at word boundaries.
An end-of-sequence token is added after each sentence.
Tokens are first converted into learned embedding, which are then passed through 3 convolution layers with dropout~\cite{srivastava2014dropout} and batch normalization layer~\cite{ioffe2015batch}.
Finally, long-term contextual information is modeled by passing the output through a single bi-directional long short-term memory (LSTM) layer with ZoneOut regularization~\cite{krueger2016zoneout}.

\subsection{Resampling}
The length of the output waveform sequence is very different from the length of encoder representations.
In Tacotron~2~\cite{shen2018natural}, this is resolved by the attention mechanism.
To make the structure non-autoregressive and speed-up  inference, we adopt the Gaussian upsampling introduced in the non-attentive Tacotron~\cite{shen2020non}.
Instead of repeating each token according to its duration, Gaussian upsampling predicts the duration and influence range simultaneously.
These parameters are used in the attention weights computation, which purely relies on the predicted position.
During training, the ground truth duration is used instead and an additional mean square loss is measured to train the duration predictor.
This is labeled as \textbf{Duration Loss} in Figure~\ref{fig:architecture}.
Ground truth duration is not needed during inference and predicted duration is adopted instead. 

\subsection{Sampling Window}
Since the waveform resolution is very high (24,000 samples per second in our case), it is not feasible to compute the loss on all waveform samples in an utterance because of the high computation cost and memory constraints.
Instead, after learning the representations on the whole input sequence, we sample a small segment to synthesize the waveform.
Due to the resampling layer, the encoder representations and waveform samples are already aligned.
Random segments are sampled individually in each minibatch and the corresponding waveform segment is extracted based on the upsampling rate (300 in our setup).
The full encoder sequence (after resampling) is used during inference which introduces a small mismatch between training and inference.
We conduct ablation studies to study how the sampling window size influences fidelity in Section~\ref{sec:exp_window}.

\subsection{Decoder}

\begin{figure}%
  \centering
  \resizebox{0.7\columnwidth}{!}{\begin{tikzpicture}[node distance=1cm, scale=1.0, every node/.style={transform shape}]
\tikzstyle{layer} = [rectangle, thick, minimum width=3cm, minimum height=0.6cm, align=center, draw=black]
\tikzstyle{block} = [rectangle, thick, rounded corners, minimum width=3cm, minimum height=0.6cm, align=center, draw=black]
\tikzstyle{block_small} = [rectangle, thick, rounded corners, minimum height=0.6cm, align=center, draw=black]
\tikzstyle{arrow} = [thick,->,-{Latex[length=2.5mm,width=2.5mm]}]
\tikzstyle{line} = [thick]

\node(i0) [align=center] at (2.75, 3.6) {$y_n$}; %
\node(i1) [layer, fill=blue!10] at (-2.75, -4.8) {Sampled Frames};

\node(o0) [align=center] at (-2.75, 4.8) {$\epsilon_n$};

\node(d0) [block, fill=yellow!20!blue!20!] at (2.75, -2.4) {DBlock (512, $/5$)};
\node(d1) [block, fill=yellow!20!blue!20!] at (2.75, -1.2) {DBlock (256, $/3$)};
\node(d2) [block, fill=yellow!20!blue!20!] at (2.75, 0) {DBlock (128, $/2$)};
\node(d3) [block, fill=yellow!20!blue!20!] at (2.75, 1.2) {DBlock (128, $/2$)};
\node(d4) [layer, fill=red!20!] at (2.75, 2.4) {$5 \times 1$ Conv (32)};

\node(u0) [layer, fill=red!20!] at (-2.75, -3.6) {$3 \times 1$ Conv (768)};
\node(u1) [block, fill=yellow!60!] at (-2.75, -2.4) {UBlock (512, $\times 5$)};
\node(u2) [block, fill=yellow!60!] at (-2.75, -1.2) {UBlock (512, $\times 5$)};
\node(u3) [block, fill=yellow!60!] at (-2.75, 0) {UBlock (256, $\times 3$)};
\node(u4) [block, fill=yellow!60!] at (-2.75, 1.2) {UBlock (128, $\times 2$)};
\node(u5) [block, fill=yellow!60!] at (-2.75, 2.4) {UBlock (128, $\times 2$)};
\node(u6) [layer, fill=red!20!] at (-2.75, 3.6) {$3 \times 1$ Conv (1)};

\node(t) at (0, 3.0) {$\sqrt{\bar\alpha}$};

\node(f0) [block_small] at (0, -2.4) {FiLM};
\node(f1) [block_small] at (0, -1.2) {FiLM};
\node(f2) [block_small] at (0, 0) {FiLM};
\node(f3) [block_small] at (0, 1.2) {FiLM};
\node(f4) [block_small] at (0, 2.4) {FiLM};

\draw[arrow] (i0.south) -- (d4.north);
\draw[arrow] (d1.south) -- (d0.north);
\draw[arrow] (d2.south) -- (d1.north);
\draw[arrow] (d3.south) -- (d2.north);
\draw[arrow] (d4.south) -- (d3.north);

\draw[arrow] (d4.west) -- (f4.east);
\draw[arrow] (d3.west) -- (f3.east);
\draw[arrow] (d2.west) -- (f2.east);
\draw[arrow] (d1.west) -- (f1.east);
\draw[arrow] (d0.west) -- (f0.east);

\draw[arrow] (f4.west) -- (u5.east);
\draw[arrow] (f3.west) -- (u4.east);
\draw[arrow] (f2.west) -- (u3.east);
\draw[arrow] (f1.west) -- (u2.east);
\draw[arrow] (f0.west) -- (u1.east);

\draw[arrow] (i1.north) -- (u0.south);
\draw[arrow] (u0.north) -- (u1.south);
\draw[arrow] (u1.north) -- (u2.south);
\draw[arrow] (u2.north) -- (u3.south);
\draw[arrow] (u3.north) -- (u4.south);
\draw[arrow] (u4.north) -- (u5.south);
\draw[arrow] (u5.north) -- (u6.south);
\draw[arrow] (u6.north) -- (o0.south);

\end{tikzpicture}}
  \caption{WaveGrad decoder. The inputs consists of the conditioning representations $x$, the noisy waveform generated from the previous iteration $y_n$, and the noise level $\sqrt{\bar{\alpha}}$. The model produces the direction $\epsilon_n$ at each iteration, which can be used to update $y_n$. FiLM is the Feature-wise Linear Modulation \cite{dumoulin2018feature-wise} which combines information from $y_n$ and $x$.}
  \label{fig:decoder}
\end{figure}
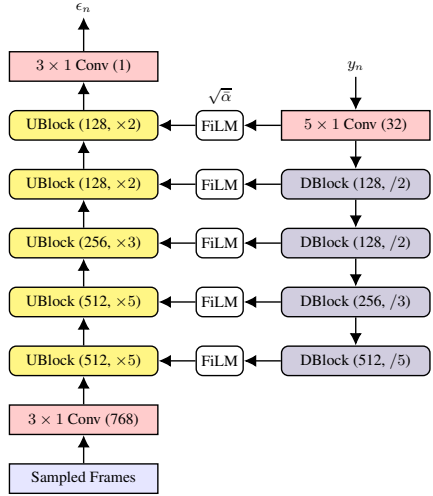

The decoder gradually upsamples the hidden representations to match the waveform resolution.
In our case, the waveform is sampled at 24 kHz and we need to upsample by 300 times.
This is achieved using the WaveGrad decoder~\cite{chen2021wavegrad}, as shown in Figure~\ref{fig:decoder}.
The architecture includes 5 upsampling blocks (UBlock) and 4 downsampling blocks (DBlock).
In each iteration of the generation process, the network denoises the noisy input waveform estimate $y_n$ by predicting the included noise term $\epsilon_n$, conditioning on the hidden presentations following equation~\ref{eq:inf}.
As described in Section~\ref{sec:scorematching}, the generation process begins from a random noise estimate $y_N$, and iteratively refines it over $N$ (typically set to 1000) steps to generate a waveform sample.
Following our previous work~\cite{chen2021wavegrad}, the training objective is the $L_1$ loss between the predicted and ground truth noise term.
During training, this loss is computed using a single, randomly sampled, iteration.

\section{Experiments}

We compare WaveGrad 2 with other neural TTS systems.
Following \cite{chen2021wavegrad}, baseline systems were trained on a proprietary dataset consisted of 385 hours of high-quality English speech from 84 professional voice talents.
A female speaker was chosen from the training dataset for evaluation.
128-dimensional mel-spectrogram features were extracted from 24 kHz waveform following the previous setup~\cite{chen2021wavegrad}: 50 ms Hanning window, 12.5 ms frame shift, 2048-point FFT, 20~Hz \& 12~kHz lower \& upper frequency cutoffs.
For Wave-Tacotron~\cite{weiss2020wave} and the proposed WaveGrad~2 models, we used a subset of training set which included all the audio of the test speaker.
This subset included 39 hours of speech.
Preliminary results suggested that WaveGrad 2 trained on a single-speaker dataset gave better performance, especially when the network size was small.

The following models were used for comparison: 

\noindent \textbf{Tacotron~2 + WaveRNN} which was conditioned on mel-spectrograms predicted by a Tacotron~2 model in teacher-forcing mode following \cite{shen2018natural}.
The WaveRNN used a single LSTM layer with 1,024 hidden units, 5 convolutional layers with 512 channels to process the mel-spectrogram features.
A 10-component mixture of logistic distributions~\cite{salimans-iclr-2017} was used as its output layer, generating 16-bit quantized audio at 24~kHz sample rate.
Preliminary results indicated that further reducing the number of LSTM units hurt performance.

\noindent \textbf{Tacotron~2 + WaveGrad} which was trained on ground truth mel-spectrograms following~\cite{chen2021wavegrad}.
Two different network sizes were included.%
The 15M-parameter WaveGrad Base model took 7,200 samples corresponding to 0.3 seconds of audio as input during training.
For the 23M-parameter WaveGrad Large model, each training sample included 60 frames corresponding to a 0.75 second audio segment.

\noindent \textbf{Tacotron~2 + GAN-TTS/MelGAN} follow the first baseline, using non-autoregressive neural vocoders.
We followed the setup and hyperparameters in their original work.
MelGAN~\cite{kumar-neurips-2019} included 3.22M parameters, trained for 4M steps while GAN-TTS included 21.4M parameters, and was trained for 1M steps.

\noindent \textbf{Wave-Tacotron}\cite{weiss2020wave} which used a Tacotron-like encoder-decoder architecture, where the autoregressive decoder network uses a normalizing flow to synthesize waveform samples directly as a sequence of non-overlapping 40ms frames.

For all baseline systems which used separate vocoder networks, we used predicted mel-spectrograms from the Tacotron~2 model during inference.
The same Tacotron~2 model was used for all baselines.

The evaluation set included 1,000 sentences.
A five-point Likert scale score (1: Bad, 2: Poor, 3: Fair, 4: Good, 5: Excellent) was adopted and the rating increment was 0.5.
Subjects rated the naturalness of each stimulus after listening to it in a quiet room.
Each subject was allowed to evaluate up to six stimuli that were randomly chosen and presented in isolation.
The subjects were native speakers of English living in the United States and were requested to use headphones.

Subjective evaluation results are summarized in Table~\ref{tab:mos}.
The WaveGrad~2 models almost matched the performance of the autoregressive Tacotron~2 + WaveRNN baseline and outperformed other baselines with non-autoregressive vocoders.

\begin{table*}[t]
\centering
\caption{ Mean opinion scores (MOS) of various models and their confidence intervals. \textbf{MT}: Multi-task learning.
}
\label{tab:mos}
\begin{tabular}{@{}lrr@{}}
 \toprule
 \bfseries Model & \bfseries Model size & \bfseries MOS ($\uparrow$) \\
 \midrule
 Tacotron~2 + WaveRNN & 38M + 18M & 4.49 $\pm$ 0.04 \\
 Tacotron~2 + WaveGrad(Base, 1000) & 38M + 15M & 4.47 $\pm$ 0.04 \\
 Tacotron~2 + WaveGrad(Large, 1000) & 38M + 23M & 4.51 $\pm$ 0.04 \\
 Tacotron~2 + MelGAN & 38M + \hskip1ex 3M &  3.95 $\pm$ 0.06 \\
 Tacotron~2 + GAN-TTS & 38M + 21M & 4.34 $\pm$ 0.04 \\
 \midrule
 Wave-Tacotron \cite{weiss2020wave} & 38M  & 4.08 $\pm$ 0.06 \\
  \midrule
  WaveGrad~2 & & \\
  \quad Encoder(2048) + WaveGrad(Large, 1000) & 193M & 4.37 $\pm$ 0.05 \\
  \quad Encoder(2048) + WaveGrad(Large, 1000) + MT & 193M & 4.39 $\pm$ 0.05 \\
  \quad Encoder(1024) + WaveGrad(Large, 1000) + MT + SpecAug & 73M & 4.43 $\pm$ 0.05 \\

 \midrule
 Ground Truth &  -- & 4.58 $\pm$ 0.05 \\
 \bottomrule
\end{tabular}
\end{table*}

\subsection{Sampling Window Size}
\label{sec:exp_window}
Memory usage is a major concern for end-to-end training.
Long sequences corresponding to multi-second utterances may not fit into memory %
since the main computation bottleneck comes from the WaveGrad decoder which operates at the waveform sample rate.
To make training efficient, we sample a small segment from the resampled encoder representation and train the decoder network using this segment instead of the full sequence.

Two different window sizes were explored: 64 and 256 frames,  corresponding to 0.8 and 3.2 seconds of speech, respectively.
The results are shown in Table~\ref{tab:window}.
The use of the large window gave better MOS compared to the small window.
In all following experiments we use the large window for training.

\begin{table}[t]
\centering
\caption{ Comparison between different sampling window sizes. All models use 1,000 iterations for inference.
}
\label{tab:window}
\begin{tabular}{@{}l@{}rr@{}}
 \toprule
 \bfseries Model & \bfseries Window Size & \bfseries MOS ($\uparrow$) \\
 \midrule
  Encoder(512) + WaveGrad(Base) & 0.8 sec & 3.80 $\pm$ 0.07 \\
  Encoder(512) + WaveGrad(Base) & 3.2 sec & 3.88 $\pm$ 0.07 \\
 \bottomrule
\end{tabular}
\end{table}

\subsection{Network Size}
We carried out ablations using different network sizes.
The encoder only needs to be computed once, thus increasing the hidden dimension has small impact to the inference speed.
On the other hand, the WaveGrad decoder needs to be executed multiple times depending on the number of iterations.

Subjective evaluation results are presented in Table~\ref{tab:size}.
It can be seen from the table that the larger encoder size increased the number of parameters by a large margin, and led to a small quality improvement.
However, the improvement was smaller compared to using a larger WaveGrad decoder, indicating that having a larger decoder is crucial.

\begin{table}[t]
\centering
\caption{ Comparison between different network sizes. All models use 1000 iterations for inference.
}
\label{tab:size}
\begin{tabular}{@{}l@{}rr@{}}
 \toprule
 \bfseries Model & \bfseries Model size & \bfseries MOS ($\uparrow$) \\
 \midrule
  Encoder(512) + WaveGrad(Base) & 37M & 3.88 $\pm$ 0.07 \\
  Encoder(512) + WaveGrad(Large)  & 40M & 4.19 $\pm$ 0.06\\
  Encoder(2048) + WaveGrad(Base)  & 188M & 4.05 $\pm$ 0.07\\
  Encoder(2048) + WaveGrad(Large) & 193M & 4.37 $\pm$ 0.05\\ 
 \bottomrule
\end{tabular}
\end{table}

\begin{table}[t]
\centering
\caption{ Impact of augmentation on the learned representations. All models use 1000 iterations for inference.
}
\label{tab:augmentation}
\begin{tabular}{@{}l@{\hskip3ex}rr@{}}
 \toprule
 \bfseries Model & \bfseries SpecAug & \bfseries MOS ($\uparrow$)\\
 \midrule
  Encoder(2048) + WaveGrad(Large)  & N & 4.37 $\pm$ 0.05 \\
  Encoder(2048) + WaveGrad(Large)  & Y & 4.40 $\pm$ 0.05 \\
 \bottomrule
\end{tabular}
\end{table}

\subsection{Hidden Features Augmentation}
We explored applying a variant of SpecAugment~\cite{park2019specaugment} to the conditioning input to the decoder (the resampled encoder output).
The augmentation is applied on the learned hidden representations instead of the spectrograms. This can be viewed as a form of correlated block dropout.
32 consecutive frames were randomly selected to be masked and we applied it twice. 
The intuition is that the WaveGrad decoder can recover the masked part by conditioning the contextual information.
This enforces the encoder to learn robust representations which include more context information. Results are shown in Table~\ref{tab:augmentation}.  We did not observe large improvements with this regularization.

\subsection{Multi-task Learning and Speed-Quality Tradeoff}
Inspired by FastSpeech~2s~\cite{ren2020fastspeech}, we explored leveraging mel-spectrogram features to enhance encoder training.
The encoder is encouraged to extract representations that can directly predict the spectrogram features.
We added a separate mel-spectrogram decoder after the resampling layer to predict the mel-spectrogram features.
This decoder included one upsampling block~\cite{chen2021wavegrad} and the mean squared error (MSE) was measured as an additional loss on the whole sequence.
During inference, we simply dropped this decoder similar to FastSpeech~2s~\cite{ren2020fastspeech}.

As shown in Table~\ref{tab:multitask}, there was no significant performance difference with multi-task training.
This suggests that multi-task learning is not beneficial for the end-to-end generation.
We also explored reducing the number of iterations from 1000 to 50 and found a small performance degradation (about 0.07 points).

\begin{table}[t]
\centering
\caption{ Impact of multi-task (MT) learning and number of iterations.
}
\label{tab:multitask}
\begin{tabular}{@{}l@{\hskip1ex}r@{\hskip2ex}r@{\hskip2ex}r@{}}
 \toprule
 \bfseries Model  & \bfseries MT & \bfseries Iter & \bfseries MOS ($\uparrow$) \\
 \midrule
  Encoder(2048) + WaveGrad(Large)  & N & 1000 & 4.37 $\pm$ 0.05  \\
  Encoder(2048) + WaveGrad(Large)  & Y & 1000 & 4.39 $\pm$ 0.05 \\
  Encoder(2048) + WaveGrad(Large)  & Y & 50 & 4.32 $\pm$ 0.05 \\
 \bottomrule
\end{tabular}
\end{table}

\section{Conclusions}
In this paper, we presented WaveGrad~2, an end-to-end non-autoregressive TTS model which takes a phoneme sequence as input and synthesizes the waveform directly without using hand-designed intermediate features (e.g., spectrograms) like most TTS systems.
Similar to prior work~\cite{chen2021wavegrad}, the output waveform is generated through an iterative refinement process beginning from random noise.
The generation procedure provides a tradeoff between fidelity and speed by varying the number of refinement steps.
Experiments demonstrate that WaveGrad~2 is capable of generating high fidelity audio, comparable to strong baselines.
Ablation studies exploring different model configurations, found that increased model size is the most important factor in determining WaveGrad~2 synthesis quality.
Future work includes improving the performance under the limited number of refinement iterations.

\bibliographystyle{IEEEtran}

\bibliography{mybib}

\end{document}